**Wafer-Bonded Surface Plasmon Waveguide Biosensors with In-Plane Microfluidic Interfaces**

Muhammad Asif,[1,2] Oleksiy Krupin,[2,3] Wei Ru Wong,[4] Zohreh Hirbodvash,[2,5] Ewa Lisicka-Skrzek,[2] Choloong Hahn,[2] R. Niall Tait,[1] Pierre Berini[2,3,5,*]

[1]Department of Electronics, Carleton University, Ottawa, Ontario K1S 5B6, Canada

[2]Center for Research in Photonics, University of Ottawa, Ottawa, Ontario K1N 6N5, Canada

[3]School of Electrical Engineering and Computer Science, University of Ottawa, Ottawa, Ontario K1N 6N5, Canada

[4]Integrated Lightwave Research Group, Department of Electrical Engineering, Faculty of Engineering, University of Malaya, 50603 Kuala Lumpur, Malaysia

[5]Department of Physics, University of Ottawa, Ottawa, Ontario K1N 6N5, Canada

*Corresponding author: berini@eecs.uottawa.ca

**Abstract**

Biosensors exploiting long-range surface plasmon polariton (LRSPP) waveguides comprised of Au stripes embedded in Cytop with integrated and encapsulated microfluidic channels are fabricated and demonstrated. A fabrication approach was devised where the lower cladding and recessed Au stripes are fabricated on a Si substrate, and the upper cladding and microfluidic channels are fabricated on a glass substrate, followed by wafer bonding to assemble the wafers into complete structures. The bond is centered over the full length of the optical path, yet no evidence of optical scattering or excess loss due to the bond could be observed, and no evidence of a bonding interface could be discerned from high-magnification cross-sectional images. We also demonstrate wafer-scale fabrication of in-plane microfluidic inlets and outlets, along with a fixture for fluidic edge coupling that provides sealed interfaces to external fluidic tubing and components. In-plane microfluidic interfaces are automatically defined along chip facets upon wafer dicing, precluding the need to drill through holes in lids. The performance of the chips was assessed by measuring the attenuation of LRSPPs on fully cladded reference waveguides, on waveguides passing through microfluidic channels, and by measuring the response of sensors to changes in refractive index produced by injecting various sensing solutions. Our fabrication approach based on wafer bonding and in-plane fluidic interfacing is compelling for low-cost high-volume manufacturing.

# 1. Introduction

An optical biosensor is an integrated analytical device that converts real-time biochemical interactions into an optical signal [1]. In general, optical biosensors exploit different optical techniques for biodetection, however, the use of surface plasmon polaritons (SPPs) is prevalent to monitor the interaction between a target analyte and a biomolecular receptor immobilised on the sensing surface.

SPPs are transverse magnetic (evanescent) surface waves that propagate along a metal-dielectric interface [2]. The electric field of SPPs is maximum at the interface and decays evanescently into the metal and the bounding medium, which makes SPPs highly sensitive to changes in the local refractive index. In a biosensing application, SPPs are excited in a prism-coupled geometry with the metal film in contact with the sensing solution on the other side of the prism. As biomaterial in solution binds to the metal surface, the effective index of the SPP changes, which changes the coupling angle (resonance condition) and the reflected power. Monitoring the reflected power over time probes the real-time interaction between the analyte in solution and the receptor molecules immobilised on the metal surface. This technique is termed surface plasmon resonance (SPR) biosensing. The first demonstration of SPR biosensing was presented in 1983 by Liedberg *et al*. [3]. Since then SPR biosensors have become a crucial technology for characterizing and measuring biomolecular interactions, given their ability to do so in a label-free, real-time manner [4].

Single-interface SPPs are strongly confined but also strongly attenuated. Conversely, long-range surface plasmon polaritons (LRSPPs) are less confined but much less attenuated [5]. The extended propagation length of LRSPPs leads to a greater overall sensitivity due to a longer optical interaction with the sensing medium [6]. LRSPPs are supported by a thin metal film or stripe surrounded by a homogeneous dielectric, and originate from the symmetric coupling of single-interface SPPs on individual metal-dielectric interfaces bounding the metal [5].

LRSPP biosensing devices have been realised in prism-coupled geometries as a thin metal film on a low-index dielectric on a prism, bounded on the other side by an aqueous sensing solution [7-10]. This approach produces high sensitivity, as demonstrated for bulk RI sensing [7] and bacteria detection [8]. In addition, the larger field penetration depth of LRSPPs can be utilized for studying the effects of toxins on cancer cells [9], or monitoring cellular micro-motion within fibroblast cells [10].

Biosensors based on LRSPP waveguides have been realised as a thin Au stripe embedded in a fluoropolymer having a refractive index close to water, with microfluidic channels etched into the cladding [11]. Once filled with an aqueous sensing medium, the microfluidic channels become optically non-invasive while supporting the propagation of LRSPPs. Such biosensors have been used for many applications, including, *e.g*., biomolecular interaction analysis [12], blood typing [13], the detection of Dengue infection in patient plasma [14,15], the detection of leukemia in patient serum [16], and the detection of bacteria in urine [17]. LRSPP waveguide biosensors offer the further advantages of compactness, integration and low cost, given that wafer-scale fabrication techniques can be applied [18]. In addition to attenuation-based sensing [11-17], phase-based sensing is also possible using integrated interferometric structures [19,20].

Most lab-on-a-chip (LOC) devices currently use top-access microfluidic interfaces, where fluidic tubing is attached to ports connected to the top of the device and holes drilled through the lid provide access to microfluidic channels [21]. In-plane (side-access) microfluidic interfacing has advantages in that holes through the lid are not required because fluidic channels become accessible directly by wafer dicing, and biomaterial deposits which occur in top-fluidic interfaces are avoided because the flow remain in-plane. In-plane microfluidic interfacing has only recently been investigated, and to a very limited extent, where all interfaces reported operate by inserting a capillary or needle into the device. Examples include insertion of compliant polymer tubing into a tapered-shaped channel [22], or insertion of a metal capillary into a channel by puncturing a hole in a PDMS plug [23]. More complex solutions include insertion of a metal capillary into a PDMS device [24], or insertion of a fluidic ferrule into the device followed by tubing insertion in the ferrule [25]. From the point of view of fabrication, incorporating additional discrete components to a LOC device can significantly increase the complexity of the fabrication process to the point where mass manufacturing becomes impossible. Furthermore, inserting thin objects (*e.g.*, needles, tubing or capillaries) into a chip requires a high spatial precision, which reduces the chances for automating the process.

In this paper, we propose and demonstrate LRSPP waveguide biosensors incorporating Au stripe waveguides with integrated and encapsulated microfluidic channels, fabricated via wafer bonding. Advantageously, wafer bonding enables the independent formation of claddings, metal stripes and fluidic channels on separate substrates, decoupling the associated process steps (top-down processing creates challenges by, *e.g.*, limiting the thermal budget available as fabrication progresses [11]). Our process yields cladded, sealed and lidded devices. We also demonstrate, for the first time, wafer-scale fabrication of in-plane microfluidic inlets and outlets, along with a fixture for fluidic edge coupling that provides sealed interfaces to external fluidic tubing and components. The device is produced entirely using wafer-scale processing, with access to the microfluidic channels provided by standard wafer dicing destined to release individual biosensor chips (through-hole drilling of a substrate is not required). Section 2 describes the device structure, Section 3 summarises the proposed fabrication process, Section 4 gives optical and biosensing results, and Section 5 gives brief concluding remarks.

## 2. Device structure and fabrication

### 2.1 Device structure

The device of interest is shown in Fig. 1, as an isometric sketch taken through the sensing region. The device incorporates Au stripe waveguides, microfluidic channels and dielectric claddings formed of the fluoropolymer Cytop [26]. The waveguides are fabricated on a Cytop lower cladding about 9 μm thick supported by a silicon substrate. The microfluidic channels are formed in the Cytop upper cladding, which is also about 9 μm thick, but supported by a separate glass substrate. A wafer bonding process is employed to produce integrated biosensing devices by bonding the wafers along the dashed line shown in Fig. 1. The Au stripes are 35 nm thick and 5 µm wide, surrounded by thick symmetrical Cytop claddings. The microfluidic channels define the sensing areas as the surfaces on which LRSPPs interact with the sensing solution. Reference waveguides remain fully embedded in Cytop to isolate them from the sensing medium and are used to verify mode excitation and optical output quality via butt-coupling (as described in detail further below). During a biosensing experiment, bio-recognition elements immobilised on the sensing surface bind with analyte in the sensing solution to produce a thin adlayer of biomaterial on the surface of the Au stripe, causing a change in the attenuation of the LRSPP and power output from the chip. The biomaterial adlayer on the Au stripe grows with time, so the optical output power also changes with time, enabling real-time monitoring of binding events.

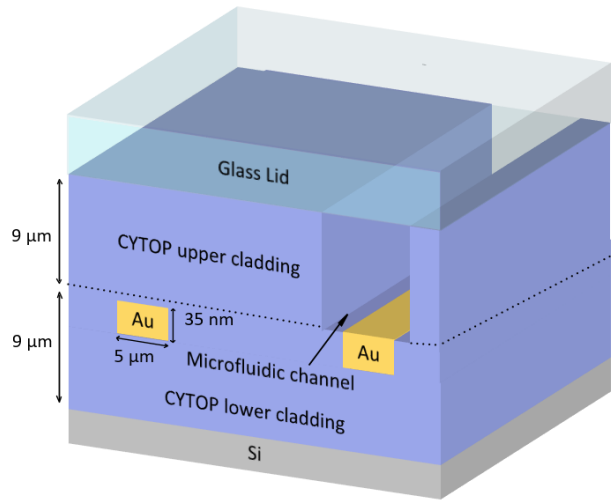

**Fig. 1.** Isometric sketch of the device of interest taken through the sensing region.

### 2.2 Fabrication process and results

Our fabrication process produces LRSPP biosensors consisting of Au stripes and microfluidic channels fabricated on separate substrates bearing the Cytop claddings, which are then joined via wafer bonding. Advantageously, this approach enables the independent formation of claddings, metal stripes and fluidic channels on separate substrates, decoupling the associated process steps. Device fabrication is depicted as the process steps illustrated in **Error! Reference source not found.**, where Parts W-1 to W-9 show the main process steps for producing waveguide structures on the lower cladding on a Si wafer, whereas Parts C-1 to C-10 depict the process steps for forming the upper cladding and microfluidic channels on a glass wafer.

A 4 inch diameter Si wafer was used as a structural base for the devices, as shown in Fig. 2 Part W-1. The wafer was silinized using APTES ((3-Aminopropyl)triethoxysilane, Sigma-Aldrich product # 440140) to promote adhesion of A-grade Cytop thereon. The lower Cytop cladding was formed by spin-coating and curing 3 layers of A-grade Cytop (CTX 809 AP2, AGC Chemicals) on the Si substrate as shown in Fig. 2 Part W-2. (Further details on the fabrication of the claddings are given in Supplementary Information, Section 1.)

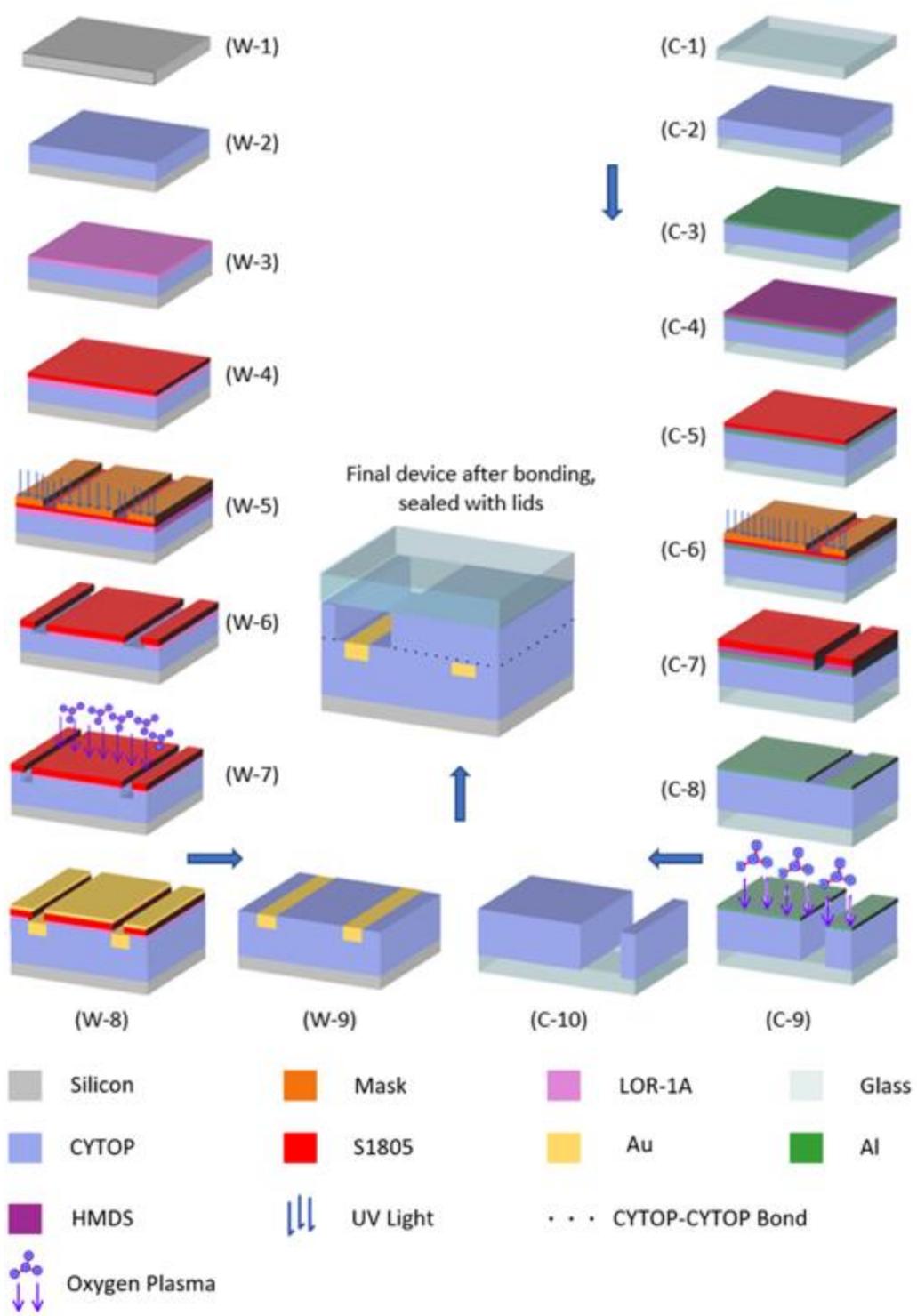

**Fig. 2.** Illustration of the process flow used for the fabrication of LRSPP biosensors. Parts W-1 to W-9 show the process steps applied to form the lower cladding and Au stripes on a Si wafer, and Parts C-1 to C-10 show the microfluidic channels formed in the upper cladding on a glass wafer.

Recessed Au stripes were realized by etching shallow trenches in the lower cladding before metal deposition following a self-aligned lithography process. The lithography stack, consisting of a lift-off resist layer (LOR 1A, MicroChem), followed by a photoresist layer (S1805, Microposit), were applied by spin-coating and curing following the manufacturer's specifications, and exposed by UV light using a photo-mask, as sketched in Fig. 2 Parts W-3 to W-5. Resist development was carried out in MF-321 (Microposit) after UV exposure, following the manufacturer's recommendations, as sketched in Fig. 2 Part W-6. A short RIE plasma etch was applied to create 35 nm deep trenches to eventually house the Au stripes, as sketched in Fig. 2 Part W-7. Fig. 3(a) shows an AFM scan of an etched trench, revealing a trench depth of 35 nm. The roughness in the trench was found to be 0.79 nm (RMS), and the roughness away from the trench (on the Cytop cladding surface), was measured to be 0.53 nm (RMS). A 35 nm thick layer of Au was then thermally deposited in an evaporator, as sketched in Fig. 2 Part W-8. Lift-off was carried out in PG-Remover (N-Methylpyrrolidinone, MicroChem) to obtain patterned Au stripes in shallow trenches, as sketched in Fig. 2 Part W-9, concluding fabrication on the lower cladding. Fig. 3(b) shows an AFM scan of a recessed Au stripe, revealing that it is about 4 nm below the surrounding Cytop surface. Such small deviations will be filled by Cytop during the subsequent wafer bonding process, producing stripes that are completely embedded. The surface roughness on the stripe is 0.85 nm (RMS), which is acceptable for LRSPP propagation thereon. (Further details on these fabrication steps are given in Supplementary Information, Section 2.)

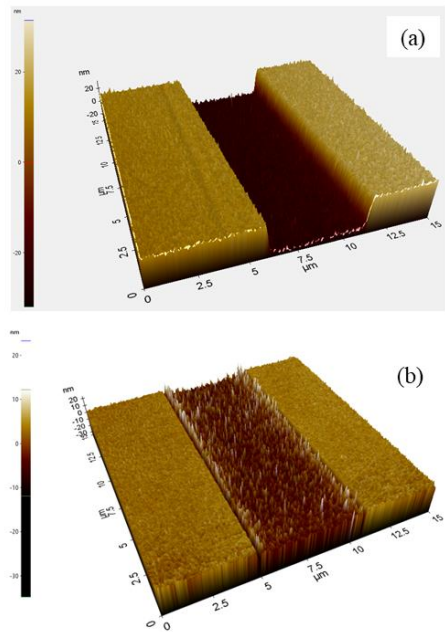

**Fig. 3.** (a) AFM scan of a trench etched into Cytop. (b) AFM scan of a recessed Au stripe deposited into a trench etched into Cytop.

A cleaned 4 inch diameter Borofloat 33 glass wafer, 300 μm thick (T24050, SIEGERT Wafers) was used as the substrate on which the upper cladding and microfluidic channels of the biosensor were formed (and eventually as the lid encapsulating the channels), as sketched in Fig. 2 Part C-

1. Borofloat glass was selected because its coefficient of thermal expansion is matched to that of Si, ensuring that stress does not build up during the thermal cycling that is necessary in the subsequent wafer bonding step. The cleaned glass wafer was silinized using APTES. The upper Cytop cladding was then formed by spin coating 3 layers of A-grade Cytop on the glass wafer, as sketched in Fig. 2 Part C-2, following the same process as applied for the lower cladding. Al was deposited by e-beam evaporation, for use as the etch mask to define the microfluidic channels, as sketched in Fig. 2 Part C-3. Fig. 2 Parts C-4 to C-8 show the lithography steps to pattern the Al etch mask on the cladding (HMDS and S1805 are applied in Fig. 2 Parts C-4 and C-5; UV exposure, development in MF-321 and resist stripping are applied in Fig. 2 Parts C-6 to C-8.) The microfluidic channels were then defined using reactive ion etching (RIE) and the Al etch mask was removed in MF321, as sketched in Parts C-9 and C-10 respectively. (Further details on these fabrication steps are given in Supplementary Information, Section 3.)

The next process step consists of wafer bonding the halves (Fig. 2 Parts W-9 and C-10) to produce cladded, sealed and lidded devices. Wafer bonding was performed in a wafer aligner and bonder, and the process controlled through the parameters of bonding temperature, bonding force and bonding time. Bonding was carried out in vacuum to avoid air traps forming at the interface, thereby producing a better bond. After aligning and contacting the wafers, and applying a force of 1000 N, heat was applied via a linear ramp of 1 °C/min. The bonding temperature was set to 115 °C, which is slightly above the glass transition temperature of Cytop ($T_g$ = 108 °C). Applying the force before heating brings the wafers into contact over their full area, and subsequent heating causes Cytop to expand which increases the bonding force, then soften and flow as $T_g$ is reached such that polymer molecules inter-diffuse over the bonding interface to form a strong bond. The wafer pair was kept under force and heat for about 16 hrs. Heating was then turned off, and the bonded pair was allowed to cool inside the chamber. Once the bonded pair reached room temperature, it was removed for inspection and further processing. Fig. 4 shows typical force and temperature curves recorded during bonding. The wafer pair required about 1.5 hours to reach the set temperature, with a slight overshoot of a few ºC. The maximum force reached was 2100 N after about 16 hrs. The increased force beyond the force initially applied (1000 N) is due to the expansion of both Cytop claddings, of total thickness of about 18 µm. Heating was turned off at about 17 hrs, when the Cytop had reached its maximum expansion (maximum force recorded). Local minima and maxima observed along the force curve may be due to material release (flow and creep) caused by the long term applied stress and heat. (Further details on the wafer bonding process steps are given in Supplementary Information, Section 4.)

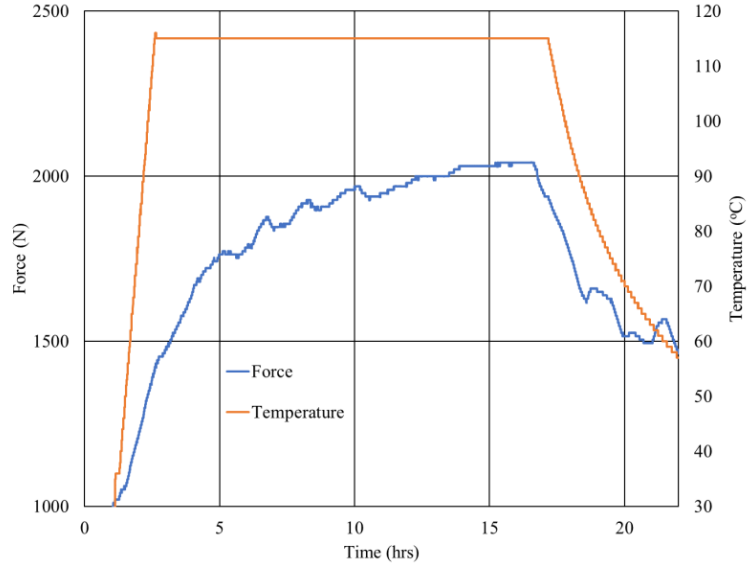

**Fig. 4.** Force and temperature evolution during wafer bonding.

Several bonded wafer pairs were inspected under optical microscope and the bonded areas were generally estimated to be greater than 90% of the total area. Fig. 5 shows a collection of microscope images of a bonded wafer pair at different locations and with different magnifications. Figs. 5(a) through 5(c) show well-bonded regions. The Au stripes are generally well-aligned in the microfluidic channels, following the layout. Fig. 5(d) shows a partially bonded region of the wafer pair - the boundary between bonded and un-bonded regions is easily observed. Un-bonded regions could be due to excessive topography or particulates at the interface of the affected regions.

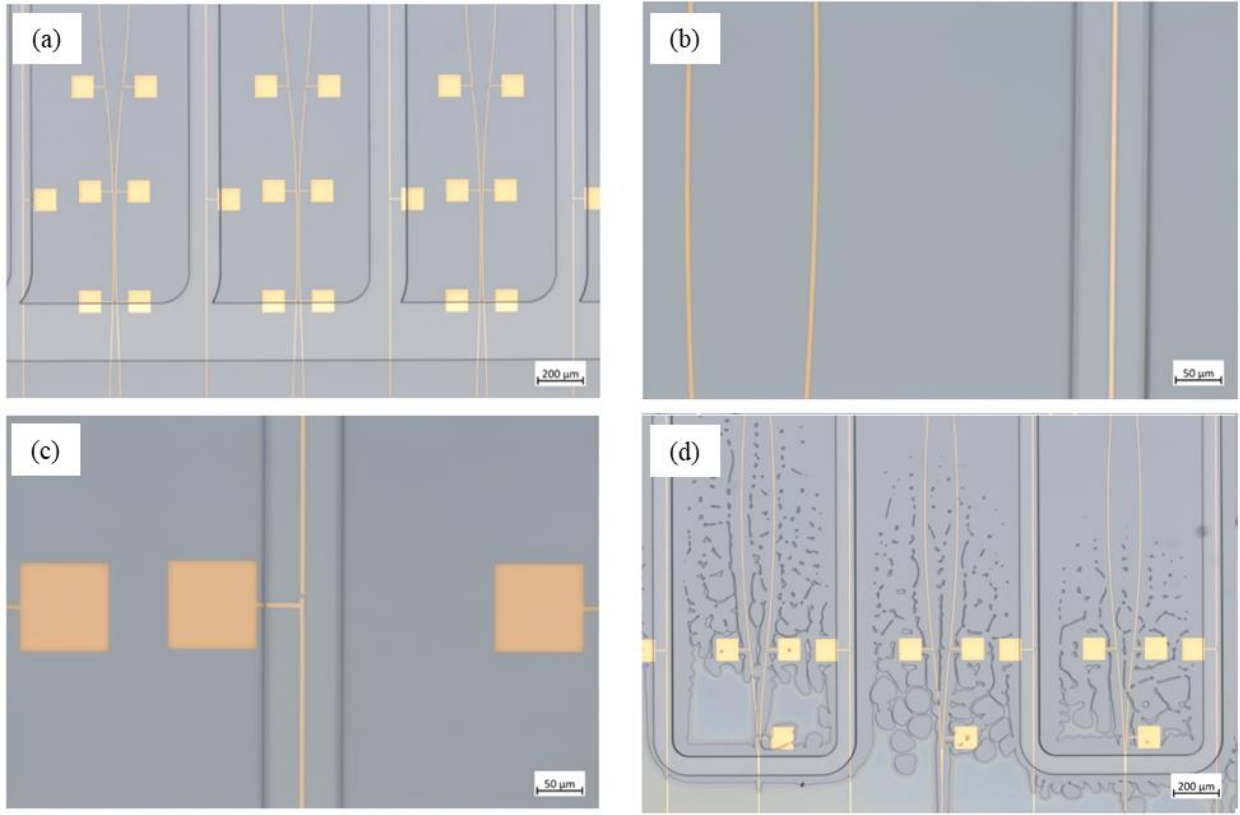

**Fig. 5.** Optical microscope images of a wafer pair post bonding; (a) – (c) well bonded regions; (d) partially bonded regions.

The bond was tested by inserting a sharp knife edge between wafers in an attempt to pry them apart, but the bonds remained intact and the wafer pairs did not de-bond. The bonds were further verified by attempting to inject IPA (isopropyl alcohol) between wafer pairs with no ingress being observed under microscope.

Several wafer pairs were diced into approximately 300 rectangular chips after completion of the bonding process. No de-bonding was observed post dicing. Fluidic inlets and outlets are opened by dicing on pairs of opposing end facets (following the layout), along the left and right edges of chips, whereas optical inputs and outputs are located on the front and back facets, as shown in Fig. 6(a).

Diced chips were randomly chosen from wafers for further inspection. IPA was injected into fluidic channels from their inlets, and flow was observed using a microscope to be well-confined to the channels, with no IPA ingression along the bonded interface being apparent. The optical and fluidic facets of a die were polished, then cleaned by ultrasonic action and dipped into a KOH bath to remove dicing and polishing debris. No de-bonding was observed after applying these processes.

Polished optical and fluidic end facets were examined using a helium ion microscope (HIM, Zeiss Orion Nanofab), as it was difficult to examine the facets under SEM due to charging. The HIM is equipped with an electron flood gun to neutralise charging by He$^+$ on insulating samples.

HIM imaging yields high-resolution, strong-contrast images with no charging artifacts. A HIM image of an optical facet is shown in the left and right facets. (b) HIM image of an optical facet. (c) HIM image of a fluidic facet.

Fig. 6(b). The Au stripe edge is visible and looks straight and intact. There are no obvious signs of stripe bending or deformation due to the bonding process. The width of the stripe is 5 µm as designed on the layout. A HIM image of a fluidic inlet is shown in

Fig. 6(c). The channel is rectangular and of high aspect ratio with no signs of deformation. The height of the channel is 9.7 µm which coincides well with the thickness of the Cytop upper cladding formed on the glass wafer. The width of this channel is 80 µm. No bonding interface is apparent form either of these cross-sectional images, even at higher magnification. Evidently, fluoropolymer has migrated across the bonding interface fusing both claddings.

**Fig. 6.** (a) Optical microscope image in top view of a diced chip showing Au stripes, and a microfluidic channel that spans the width of the chip and aligned with several straight stripes. The optical inputs / outputs are taken through the front and back facets of the chip. The fluidic inlets / outlets are located on

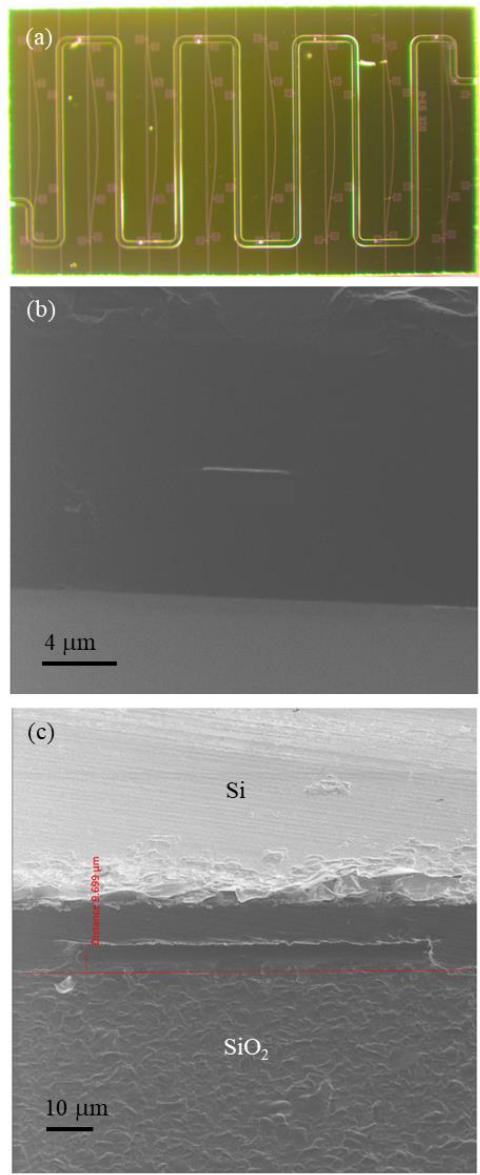

## 3. Optical and sensing measurements

### 3.1 Materials

2-Isopropanol (IPA, 733458), acetone (270725), glycerol (49767), and phosphate buffered saline (PBS, P5368) 0.01 M, pH 7.4 were obtained from Sigma-Aldrich. PBS solution was prepared by dissolving packaged salts in 1 L of distilled/deionized water (DDI $H_2O$).

### 3.2 Sensing set-up

Several chips were chosen from our diced wafers and tested optically and for sensing performance. Cleaning procedures were developed for cladded and fluidic measurements and applied prior to testing. A chip was washed in acetone to remove dicing resist (SPR-220), then placed in an ultrasonic acetone bath for 5 min to clean the optical and fluidic facets from dicing debris. This was followed by washing with IPA and drying under N2. Next, the chip was placed in a UV/Ozone cleaning chamber (Novascan, PDS) for 15 min lamp-on and 15 min lamp-off to remove possible organic contaminants from the Au waveguide surface accessible via the microfluidic inlets and outlets.

Fig. 7 illustrates our test setup. The optical path originates from a polarised DFB (distributed feedback) laser diode operating at the free-space wavelength of $\lambda_0 = 1310$ nm (NLK1356STG, NTT Electronics), and terminated with a polarisation-maintaining (PM) single-mode optical fibre (PMJ-3AX-1300-7/125-1-1-1, OZ Optics). The optical fibre is polarisation-aligned (TM) to the input of a waveguide on chip (butt-coupled), and light is injected through its front facet to excite LRSPPs propagating along a Au stripe. Output light emerges from the back facet, is collimated then split for detection by a power sensor and visualisation by an infrared camera. A variable aperture is inserted in the output path to remove background (unguided) light from detection.

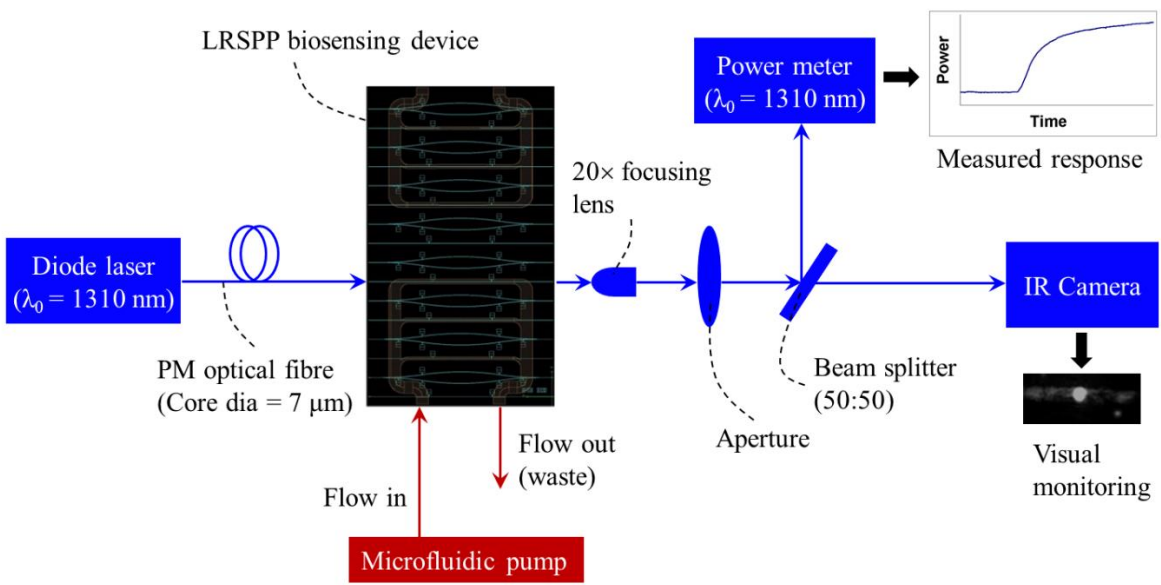

**Fig. 7.** Schematic representation of the optical biosensor interrogation setup.

The set-up comprises a microfluidic syringe system (SPM - Zero Dead Volume Diluter, Advanced MicroFluidics, Switzerland) containing eight fluidic ports with a 50 μL syringe, and a custom fluidic jig to mount the chip and provide interfacing to the fluidic facets on the left and right edges of a chip. The jig is precision-machined from several blocks of Al (Zandbelt Machine Works) as shown in Fig. 8. It consists of the following components: a metal base with a pedestal which supports the chip and two clamping arms (Fig. 8(a)). Each clamping arm has two through holes 600 μm in diameter to attach fluidic tubing and one through hole at the bottom to accommodate a clamping screw. PEEK tubing (360 μm outer diameter, 50 μm inner diameter, IDEXX Laboratories) was threaded and glued (epoxy) into the fluidic through holes of the clamping arms (Fig. 8(c)). The fluidic seal between the metal clamping arms and the fluidic facets on chip is provided by polydimethylsiloxane (PDMS) gaskets ~300 μm thick (fabricated by National Oceanography Centre, Southampton, UK) attached with epoxy to the side of each arm (Fig. 8(c)). Assembly is performed by placing the biosensor chip onto the pedestal, and sliding both clamping arms along the guiding pins which are permanently attached to the metal base, towards the chip until a good fluidic seal is achieved (Fig. 8(a), yellow arrows). The whole assembly is then tightened with clamping screws (Fig. 8(b)). This jig design can support various fluidic architectures, such as U-shape channels on the left and right sides of a chip, where the fluidic inlet and outlet are on the same facet - such a design enables two independent bioassays to be performed at the same time on the same chip. Alternatively, channels can be routed from an inlet on one side of a chip to an outlet on the other side. (Examples of both fluidic architectures are discussed below.)

An advantage of such a system is the possibility to mass produce LOC devices which can be easily integrated with external fluidics (potentially automatically) without any post-fabrication modification of devices. Although the cross-section of our microfluidic channels is small ($100 \times 10$ μm$^2$), due to the large hole in the PDMS gasket (~600 μm dia.), the lateral tolerance for microfluidic alignment is ±250 μm and the vertical tolerance is ±290 μm, which is easy to respect. Moreover, our microfluidic jig design can accept all types of channel cross-sections as long as their dimension are smaller than the hole diameter of the PDMS gasket.

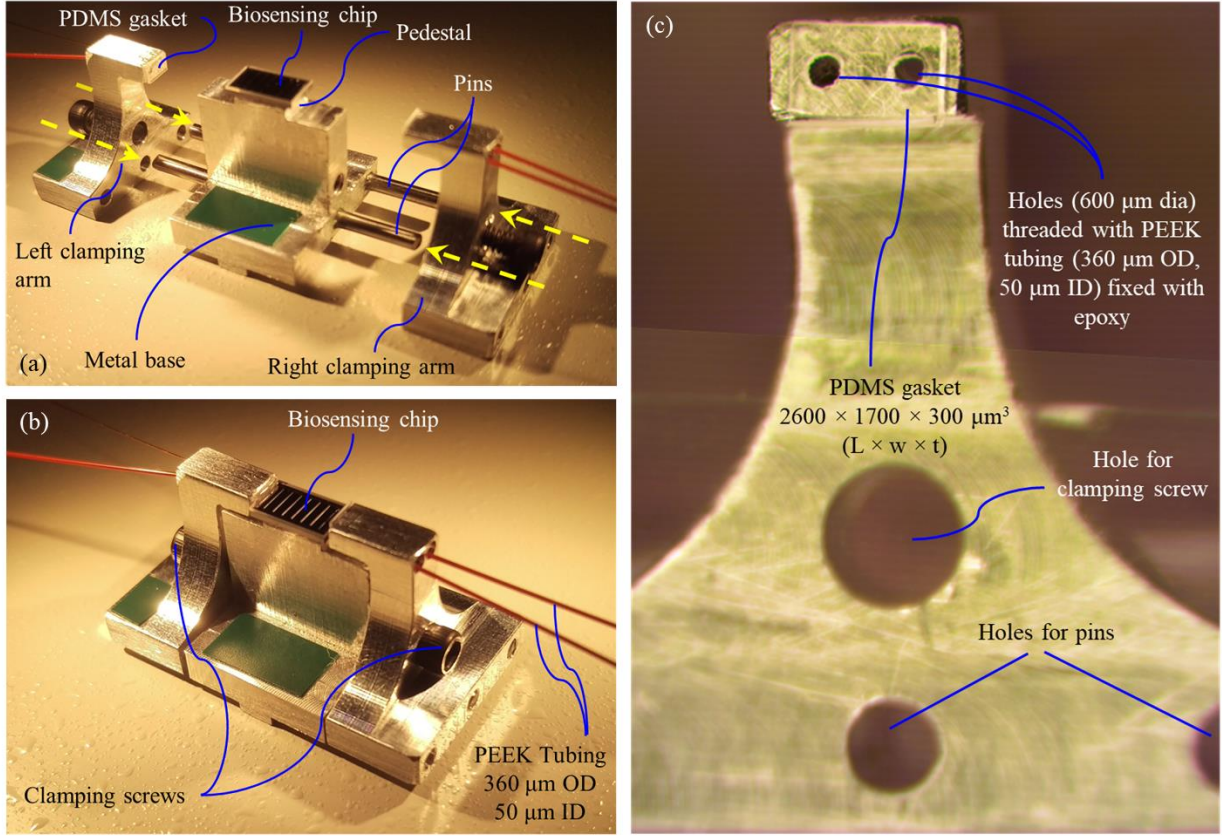

**Fig. 8.** Custom fixture designed to house a biosensor chip and provide fluidic interfaces.

### 3.3 Optical cut-back measurements

The insertion loss of a straight waveguide passing through a fluidic section, as shown in Fig. 9(a), is defined as $IL = P_{inc} - P_{out}$ [dB] and given by [dB]:

$$IL = C_C + MPA_C(L_D - L_F) + 2C_F + MPA_F L_F \qquad (1)$$

where $P_{inc}$ and $P_{out}$ [dBm] are in the incident and output optical powers, respectively, $C_C$ and $C_F$ are the coupling losses [dB] at the input facet of the chip and at a cladded-fluidic waveguide interface, respectively, $MPA_C$ and $MPA_F$ are the mode power attenuation [dB/mm] in the cladded and fluidic portions of the device, $L_D$ [mm] is the total length of the device, and $L_F$ [mm] is the length of the fluidic channel.

For entirely cladded devices (no fluidic channel) $L_F = 0$, $C_F = 0$ and Eq. (1) simplifies to $IL = C_C + MPA_C \cdot L_D$. Thus, a plot of the measured *IL vs.* $L_D$ will reveal the input optical coupling loss $C_C$ and attenuation $MPA_C$ as the intercept and slope of a best-fit linear model through the data. Both of these quantities depend strongly on fabrication quality. Three cladded waveguide lengths of

$L_D$ = 3.0, 3.8 and 4.8 mm were available. Sets were selected from four different wafers and cut-back measurements and mode outputs were obtained for each. Fig. 9(b) summarises the measurements and gives the mode outputs from the waveguides on wafer MA-41 in inset. The $MPA_C$ of the LRSPP on these wafers varies from 3.7 to 4.9 dB/mm and the measured coupling losses ($C_C$) are in the range of 0.6 to 2 dB. The attenuation ($MPA_C$) is lower than expected for a 35 nm thick 5 μm wide Au stripe embedded in Cytop, which should be about 7 dB/mm [27]. The lower attenuation may be caused by slight compression of the Au stripes due to the pressure applied during wafer bonding - even a slight change of 1 to 3 nm in Au thickness would suffice to reduce the attenuation measurably (Au is ductile) [27]. This hypothesis is supported by mode outputs that appear slightly larger than expected when visualised on the IR camera - thinner Au stripes provide lower confinement and have larger modes.

The lower than expected attenuation and the high quality of the mode outputs suggest that fabrication is generally of good quality. The roughness in the trenches and on the Au stripes (< 1 nm) have a negligible effect on the attenuation. The bonded interface produces no visual or quantitative evidence of scattering, despite being centered in the optical (LRSPP) path over the entire length of the waveguides. The lack of scattering suggests that there are no gaps along the bond, and that Cytop has flowed sufficiently across the interface to fuse the upper and lower claddings and adhere to the top Au surface. However, the compression of Au stripes due to the force applied during bonding is uncontrolled, leading to non-uniform waveguide performance as highlighted by the measured spread in $MPA_C$ and $C_C$. Optimisation of the bonding process would be required to reduce the spread.

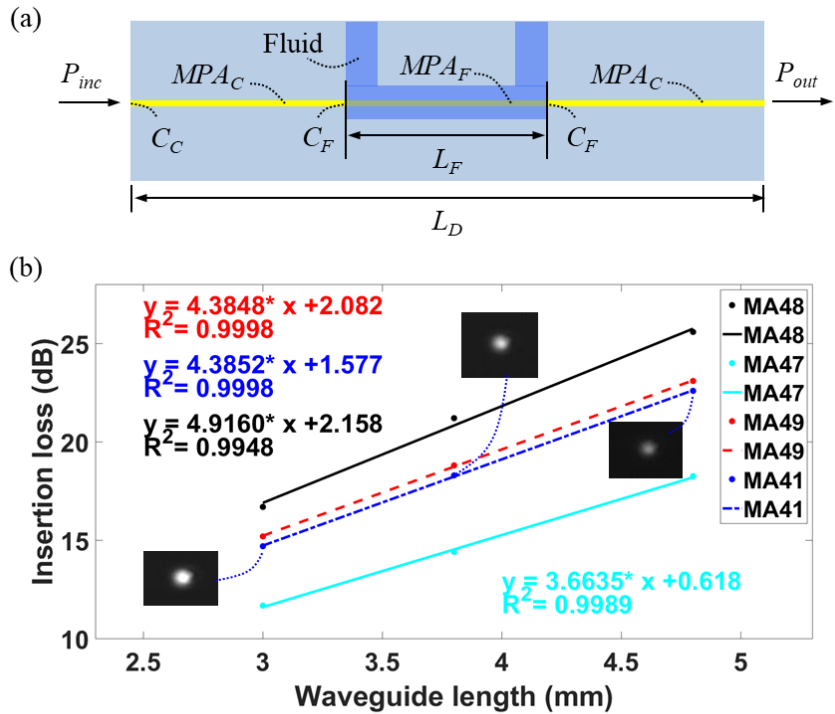

**Fig. 9**. (a) Sketch of a straight waveguide passing through a fluidic section. (b) Cutback measurements obtained at $\lambda_0 = 1310$ nm for completely cladded wafer-bonded waveguides, using three chip lengths of $L_D$ = 3.0, 3.8 and 4.8 mm, taken from four wafers (MA41, MA47, MA48, MA49). Example mode outputs are given in inset for the three waveguides taken from wafer MA41.

A fluidic chip was designed specifically to determine the LRSPP attenuation within fluidic channels. Waveguides in fluidic channels of several lengths were designed on the same chip, as shown in the layout image of Fig. 10(a). The chip of length $L_D = 4.8$ mm has 20 Au stripe waveguides arranged in pairs within different segments of a single serpentine microfluidic channel 70 μm wide. Even-numbered waveguides (Fig. 10(a)) pass through fluidic channels of different length $L_F$ in the range of 0.95 to 3.617 mm, and all odd-numbered waveguides pass through fluidic channels of the same length fixed to $L_F = 70$ μm.

As noted earlier, the results from our cladded cutback measurements reveal a spread in waveguide performance. To avoid chip-to-chip variations, we re-arrange Eq. (1) as follows:

$$IL - [C_C + MPA_C L_D] = 2C_F + L_F(MPA_F - MPA_C) \qquad (2)$$

where $C_C + MPA_C L_D$ is the insertion loss of a fully-cladded waveguide of length $L_D$. All odd-numbered waveguides have a very short fluidic channel of length $L_F = 70$ μm, so we approximate them as fully-cladded and use them as on-chip references; *i.e.*, their measured insertion loss is averaged and taken as the term $C_C + MPA_C L_D$ in Eq. (2).

We measured the insertion loss of all waveguides on a chip as a function of the refractive index of the solution injected into the fluidic channel. Three solutions consisting of mixtures of PBS with glycerol were prepared, such that they have refractive indices of $n = 1.335$ (matched to Cytop A), $n = 1.330$ (below Cytop) and $n = 1.337$ (above Cytop). A plot of $IL - [C_C + MPA_C L_D]$ *vs*. $L_F$ is linear for each solution, and the best fitting linear model for each has a slope of $MPA_F - MPA_C$ and an intercept of $2C_F$. Thus, the slope gives the increase in mode power attenuation of waveguides in fluidic channels over cladded ones, and the intercept gives the butt-coupling losses at interfaces with the fluidic channels.

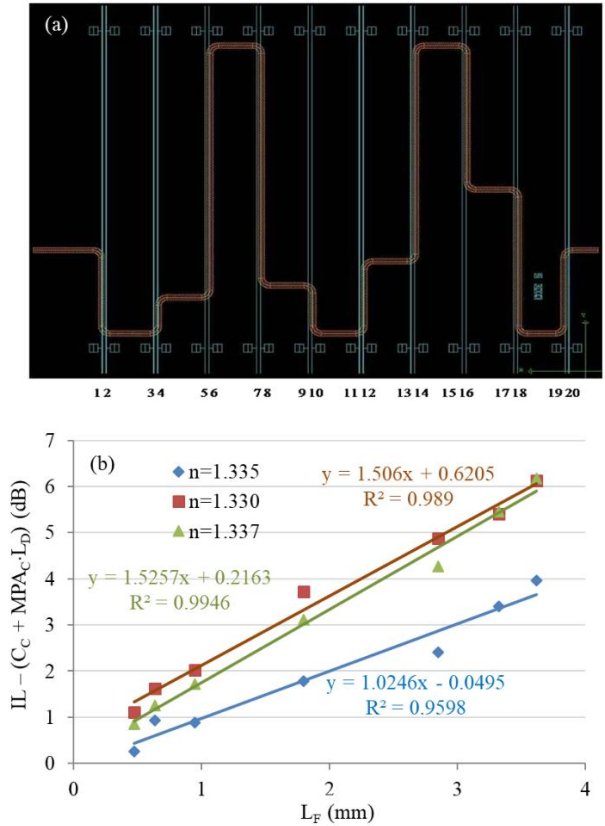

**Fig. 10.** Cut-back of waveguides in microfluidic channels. (a) Layout image of a fluidic cut-back chip of length $L_D = 4.8$ mm and varying microfluidic channel length $L_F$. (b) Plot of $\{IL - [C_C + MPA_C L_D]\}$ vs. $L_F$ for waveguides taken from a chip originating from wafer MA41.

The measured results for the three solutions are plotted in Fig. 10(b). For the case where the Cytop matching solution was injected ($n = 1.335$), an increase of 1 dB/mm in LRSPP mode power attenuation is obtained for waveguides in fluidic channels relative to the cladded waveguides on chip. The intercept in this case is -0.05 dB which is within experimental error and implies that there is no significant coupling loss at cladded-to-fluidic waveguide transitions. This negligible coupling loss and the modest increase in MPA suggest high quality fabrication of the waveguides and microfluidic channels. When solutions with $n = 1.330$ and $n = 1.337$ are injected, higher slopes of ~1.5 dB/mm and higher cladded-to-fluidic coupling losses are measured. This is expected because the MPA and coupling loss of the LRSPP in a microfluidic channel increase with asymmetry, as caused here by a refractive index mismatch between the top cladding (*i.e.*, fluid) and the bottom cladding (Cytop). Indeed, induced asymmetry is exploited in LRSPP bulk and surface attenuation-based sensing [11].

### 3.4 Bulk sensing

Five solutions of different refractive index were prepared as a mixture of PBS and glycerol, such that an index increment of $10^{-3}$ was produced between them. The solutions were injected sequentially into the system at a flowrate of 4 µL/min. A biosensor chip of the architecture shown in the layout image of Fig. 11(a) was used. Two U-shaped microfluidic channels with their inlet and outlet on the same facet are designed to isolate the left half from the right half of the chip, thereby enabling independent fluids to be used on the same chip. Such a design enables two independent bioassays to be performed at the same time on the same chip. Our fixture provides independent fluidic interfacing to each facet (Fig. 8(a)).

Fig. 11(b) gives the measured response for the biosensor of Fig. 11(a). The solutions were allowed to flow for 5 min before exchange. The highest output power of -21 dBm was obtained for the solution having $n = 1.335$, which coincides with the refractive index of Cytop (this solution was included in the set used for the fluidic cutback measurements of Fig. 10(b)). The largest signal change of $\Delta S = 1.96$ µW was obtained for the index step from $n = 1.332$ to $n = 1.333$. The standard deviation of the measured power signal, taken over a 5 min time interval (at $n = 1.333$, first cycle), was calculated to be $\delta = 4$ nW which produces a signal-to-noise ratio of $\Delta S/\delta = 490$ and a limit of detection (LOD) of $2 \times 10^{-6}$ RIU. These results are comparable with our previous measurements using open biosensors (non-encapsulated channels) [11,27]. Mode outputs were captured using frame-grabber software (LBA-710PC, Ophir Spiricon) and are also shown as insets to Fig. 11(b).

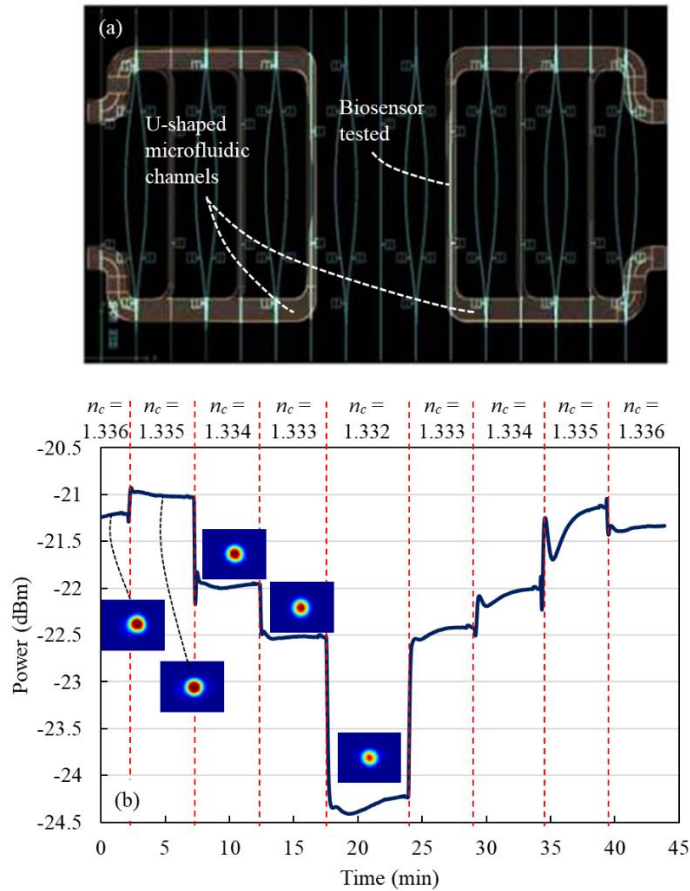

**Fig. 11.** (a) Layout image of the tested biosensor having U-shaped microfluidic channels ($L_D = 3.8$ mm). (b) Optical response of a biosensor as PBS/Glycerol solutions of different refractive index are injected

under a continuous flowrate of 4 μL /min, at $\lambda_0$ = 1310 nm and $P_{inc}$ = 5.0 dBm. The solutions vary in refractive index in increments of $1\times10^{-3}$. The cycle is repeated once.

## 4. Conclusions

Biosensors based on LRSPP waveguides incorporating a Au stripe embedded in Cytop with integrated and encapsulated microfluidic channels were proposed and demonstrated. A fabrication approach was devised where the lower cladding and recessed Au stripes were fabricated on a Si substrate, and the upper cladding and microfluidic channels were fabricated on a glass substrate. This approach has as an advantage the independent formation of the claddings, enabling appropriate thermal treatments to be applied in curing, decoupled from previously completed fabrication steps. Wafer bonding was then used to assemble the wafers into a full structure. A bonding temperature slightly above $T_g$ of Cytop was applied, along with sufficient force to ensure polymer migration and a strong Cytop-Cytop fusion bond. No evidence of a bonding interface could be observed from cross-sectional images taken at high magnification. The bond is centered over the full length of the optical path, yet no evidence of optical scattering or attenuation due to the bond could be observed.

The devices were produced entirely using wafer-scale processing, with access to the microfluidic channels provided by inlets and outlets defined automatically along left and right chip facets upon wafer dicing, with optical inputs and outputs defined along the front and back chip facets. A custom fixture for fluidic edge coupling, that provides sealed interfaces to external fluidic tubing and components, was proposed and demonstrated. In-plane microfluidic interfacing has advantages in that holes through the lid are not required and biomaterial deposits, which may occur in top-fluidic interfaces, are avoided because the flow remains in-plane.

LRSPPs were excited by butt-coupling a PM single-mode optical fiber to the optical input of a chip. LRSPP losses on Cytop cladded waveguides were measured using a cutback technique for all of the wafers produced, with MPAs ranging from 3.7 to 4.9 dB/mm. This attenuation is slightly lower than expected, suggesting slight flattening of Au stripes during wafer bonding. LRSPP losses within microfluidic channels were found to be 1 dB/mm higher compared to Cytop cladded waveguides. Bulk sensing was performed by injecting 5 solutions of varying refractive index in increments of $10^{-3}$, resulting in detection steps having a signal-to-noise ratio of 490, and yielding a LOD of $2\times10^{-6}$ RIU. Overall, the biosensors demonstrated satisfactory sensing capabilities, however modifications in the cross-sectional dimensions of the channels (specifically, increasing their height) will reduce the pressure drop required to drive the fluid.

**Supplementary information for: Wafer-Bonded Surface Plasmon Waveguide Biosensors with In-Plane Microfluidic Interfaces**

Muhammad Asif,[1,2] Oleksiy Krupin,[2,3] Wei Ru Wong,[4] Zhoreh Hirbodvash,[2,5] Ewa Lisicka-Skrzek,[2] Choloong Hahn,[2,4] R. Niall Tait,[1] Pierre Berini[2,3,5,*]

[1]Department of Electronics, Carleton University, Ottawa, Ontario K1S 5B6, Canada
[2]Center for Research in Photonics, University of Ottawa, Ottawa, Ontario K1N 6N5, Canada
[3]School of Electrical Engineering and Computer Science, University of Ottawa, Ottawa, Ontario K1N 6N5, Canada
[4]Integrated Lightwave Research Group, Department of Electrical Engineering, Faculty of Engineering, University of Malaya, 50603 Kuala Lumpur, Malaysia
[5]Department of Physics, University of Ottawa, Ottawa, Ontario K1N 6N5, Canada
*Corresponding author: berini@eecs.uottawa.ca

## 1. Fabrication of claddings

The starting wafer (Si or Borofloat 33) was thoroughly blown with $N_2$ to remove dust particles to avoid uneven spreading of Cytop and defects in the cladding. The presence of particles may also cause de-bonding of the cladding. The wafer was treated with APTES ((3-Aminopropyl)triethoxysilane, Sigma-Aldrich product # 440140) to promote adhesion of A-grade Cytop thereon (CTX 809 AP2, AGC Chemicals). The Cytop claddings were formed by spin-coating and curing 3 layers of A-grade Cytop on the substrate.

Cytop was poured by hand directly in the middle of the wafer. A settling time of 1-2 min was allowed to lapse before starting the spin coater. A rest time of 5 min at room temperature was allowed to lapse after coating each layer, followed by a 10 min soft bake at 50 °C on a hotplate. The soft bake allows the layer to de-gas and for solvent present in the Cytop to evaporate. Finally, the wafer was baked at 200 °C for 4 hrs using temperature ramps, to completely remove solvent from the layer and promote its adhesion to the APTES-coated Si substrate. The cladding process is summarised in Table 1.

**Table 1**: Cytop cladding application process.

| Activity | Description |
| --- | --- |
| 1st layer of CYTOP | Pour CYTOP A-grade in middle of the wafer, allow 1-2 min to de-gas, spin-coat at 550 rpm for 10 s followed by 1500 rpm for 30 s. |
| Soft bake | Allow 5 min at room temperature (RT) before baking on a hotplate at 50 °C for 10 min. |
| 2nd and 3rd layer of CYTOP | Repeat previous two steps. |
| Final hard bake | Ramp from RT to 80 °C, hold for 1 hr, ramp to 150 °C and hold for 1 hr, ramp to 200 °C and hold for 4 hr, then ramp down to RT. The ramp rate was set to 100 C/hr for all steps. |

## 2. Fabrication of recessed Au stripes

The Au stripes were fabricated in shallow trenches to produce recessed structures. This provides a planar surface suitable for wafer bonding and protects the stripes from tearing and deformation. After completion of the bi-layer lithography and development process on the lower Cytop cladding (Fig. 2, Part W-6, main text), shallow trenches were etched (Fig. 2, Part W-7, main text) to house the 35 nm thick Au stripes. The RIE etch parameters were experimentally determined to produce uniform and smooth trenches. Trenches were etched for 17 s in a RIE system (MARCH Juniper II) using 50 W of power and a 150 sccm $O_2$ flow rate. AFM and SEM images of a trench

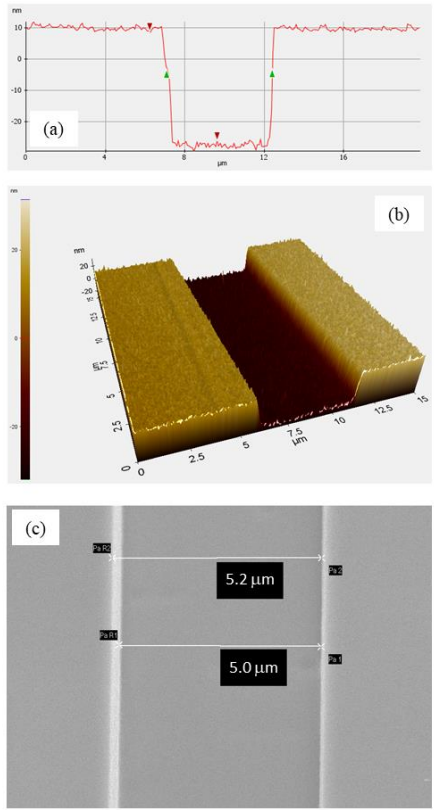

are given in

Fig. **S1**. The AFM line scans (Figs. S1(a) and S1(b)) show the depth of the trench to be 35 nm. The roughness in the trench was found to be 0.79 nm (RMS) and the roughness away from the trench (on the Cytop cladding surface), was measured to be 0.53 nm (RMS).

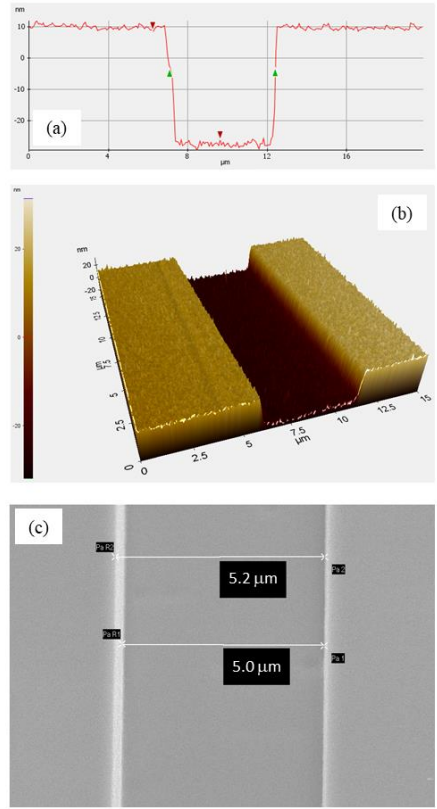

**Fig. S1.** (a) AFM line scan of a trench profile; (b) 3D AFM scan of a trench; (c) SEM image of a trench.

After trench etching and without removing the photoresist etch mask, Au was thermally evaporated, and a standard lift-off technique was used to form the stripes in the trenches (Fig. 2, Parts W-8 and W-9, main text). Fig. S2 shows AFM scans of a recessed Au stripe, revealing that it is about 4 nm below the surrounding Cytop surface. Such small deviations will be filled by Cytop during the subsequent wafer bonding process, producing stripes that are completely embedded.

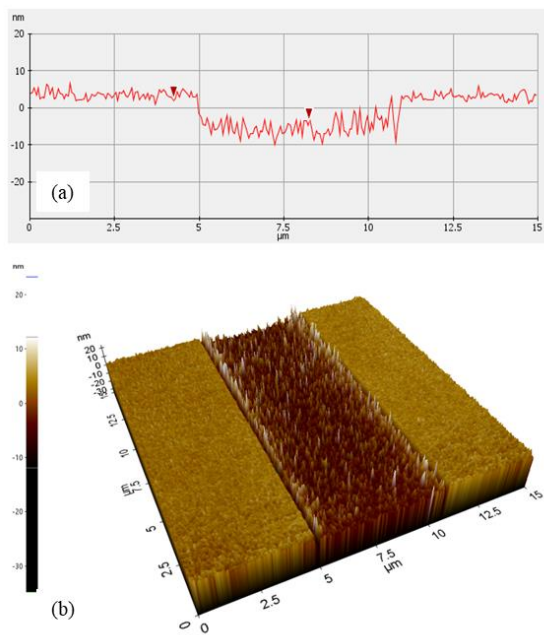

**Fig. S2.** AFM scans of a recessed Au stripe in a trench etched into Cytop. (a) Line scan; (b) 3D view.

### 3. Fabrication of microfluidic channels

Microfluidic channels were formed by etching the 9 μm thick Cytop upper cladding fabricated on the glass substrate (Fig. 2, Part C-9, main text). The Cytop cladding was etched to the glass substrate, with the substrate acting as the etch stop, and eventually the lid encapsulating the channels. These are advantages of fabricating the microfluidic channels on a separate glass substrate. It is desirable for the channel sidewalls and substrate to be as smooth as possible to avoid dead volumes, the accumulation of biomaterial, and the possible cross-contamination between sensing fluids. To minimise sidewall and substrate roughness, RIE etching was characterized through experiments, and optimal parameters were obtained. AFM and SEM imaging was performed on samples to verify the sidewall and substrate surface quality. Thus, 100 W of RF power and a 250 sccm $O_2$ flow rate were used for channel etching.

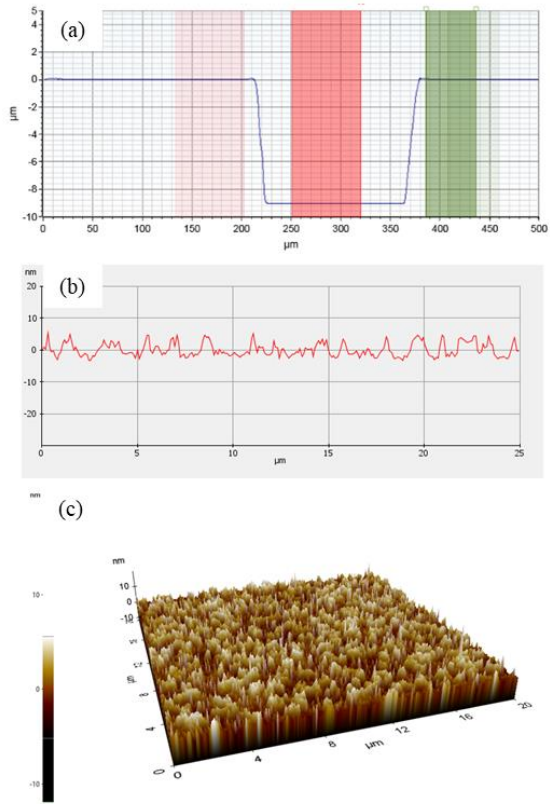

**Fig. S3**S3 shows profilometer and AFM scans of a microfluidic channel. The depth of channels was measured at several locations and found to be about 9 µm in accordance with the Cytop cladding thickness. The roughness of the substrate post etching was found to be 1.6 nm (RMS) - this region of the substrate becomes the lid of the microfluidic channels after wafer bonding.

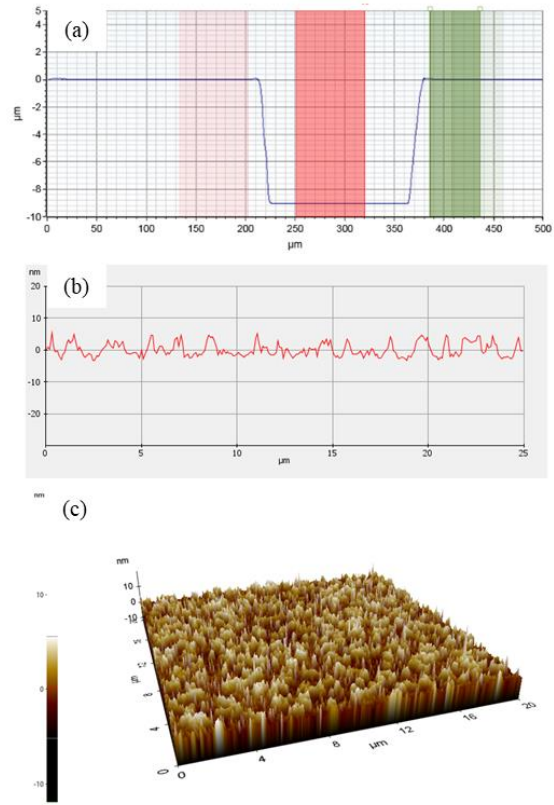

**Fig. S3.** Microfluidic channel after etching. (a) Channel depth measured using a profilometer; (b) AFM line and (c) 3D scans of channel substrate.

## 4. Wafer bonding

Before wafer bonding, the biosensing devices exist on two separate substrates: recessed Au stripes on the lower cladding fabricated on a silicon wafer, and microfluidic channels etched into the upper cladding fabricated on a glass wafer. The wafer bonding was performed in a wafer aligner and bonder, and the process controlled through parameters such as the bonding temperature, bonding force and bonding time. In addition, the vacuum pressure can be controlled in the bonder chamber to avoid air traps forming at the interface, thereby producing a better bond. The accuracy of the bonding force and pressure applied also depend on the chamber vacuum pressure.

The wafers were blown with $N_2$ before loading them into the wafer bonder. The glass substrate bearing microfluidic channels in the upper cladding was placed on the upper platen and mounted using a spring clamp with a plunger adjusted such that the wafer was forced into a slightly convex shape (wafer center protruding slightly) to ensure that first contact would occur at the central location of the wafer pair. The silicon wafer bearing the Au stripes on the lower cladding was placed flat on the lower platen. The chamber was closed and purged with $N_2$, then evacuated to a pressure of the order of $10^{-5}$ mbar. The wafers were then brought close to each other with about 2 mm of separation between them. The wafers were aligned, contacted and the force was applied. Ramped heating was turned on after the bonding force was applied. The initial force applied increased during bonding due to the thermal expansion of Cytop. The applied temperature and force caused the Cytop to flow slightly, ensuring the inter-diffusion of polymer chains across the bonding interface and leading to strong bonding. The resulting bond may achieve a strength similar to bulk Cytop. The wafer pair was kept under force and heat for about 16 hrs. Heating was then turned off, and the bonded pair was allowed to cool inside the chamber. Once the bonded pair reached room temperature, it was removed for inspection and further processing.

The wafers were aligned inside the wafer bonder chamber. The upper platen has two holes with magnification and a viewing camera for alignment. The wafers were aligned using visible dicing marks, microfluidic channel edges and Au stripe features. The wafers were brought to a 2 mm separation to start the alignment process. Initial alignment was performed at this distance. Then the separation was decreased to 1 mm, and further alignment was performed. This procedure was repeated further at separations of 0.75 and 0.25 mm. Placing the wafers closer during alignment could cause them to contact and stick inadvertently due to topography, and the alignment process would need to be repeated. Upon satisfactory alignment, the wafers were contacted, and a force of 1000 N was applied. Given that the glass wafer in the upper platen was forced into a slightly convex shape, contact was initiated near the centre of both wafers and the bond propagated radially outward as force was applied.

After aligning and contacting the wafers, and applying a force of 1000 N, heat was applied via a linear ramp of 1 °C/min. The bonding temperature was set to 115 °C, which is slightly above the glass transition temperature of Cytop ($T_g$ = 108 °C). Applying the force before heating brings the wafers into contact over their full area, and subsequent heating causes Cytop to expand which increases the bonding force, then soften and flow as $T_g$ is reached such that polymer molecules diffuse over the bonding interface to form a strong bond (applying heat before applying force can cause misalignment and deformation of channels). The wafer pair was kept under force and heat for about 16 hrs. Heating was then turned off, and the bonded pair was allowed to cool inside the chamber. Once the bonded pair reached room temperature, it was removed for inspection and further processing. The force and temperature curves recorded during a typical bonding process are reported in Fig. 4 of the main text.